\documentclass[runningheads]{svmult}

\usepackage{makeidx}   
\usepackage{graphicx}  
\usepackage{subeqnar}  
\usepackage{multicol}  
\usepackage{physprbb}  
\makeindex             

\begin{document}
\title*{Probing the gravitational redshift effect from the relativistic jets of compact AGN}
\toctitle{Focusing of a Parallel Beam to Form a Point
\protect\newline in the Particle Deflection Plane}
%
%
\titlerunning{Gravitational redshift from the jets of AGN}
%
\author{Tigran G. Arshakian\inst{1}\fnmsep\thanks{On leave from
           Byurakan Astrophysical Observatory, Byurakan 378433,
           Armenia}}
%
\authorrunning{Tigran Arshakian}
%
%
\institute{Max-Planck-Institut f\"ur Radioastronomie, Auf
              dem H\"ugel 69, 53121 Bonn, Germany} 


\maketitle              

\begin{abstract}
I explore a possibility to measure the gravitational redshift (GR)
effect in the gravitational field of massive central nuclei residing
in active galaxies (AGN). The activity of central nuclei is associated
with the bipolar jet ejection of relativistic plasma which produces 
strong radio emission. I consider the behavior of the flux density
variations of the jet plasma as a result of GR effect from a  
Schwarzschild black hole, and I discuss
possibilities to detect the GR effect from the relativistic jets
of compact AGN with present and future radio facilities.
\end{abstract}

\section{Introduction}
Development of new astronomical facilities provides high resolution
and sensitivity thus allowing one to reach the scales where the
General Relativity effects can be tested. There is already compelling
evidence that active galaxies host super massive black holes (SMBH)
reaching up to $\sim 10^{10}$ solar masses. The central active nuclei are
associated with the relativistic jets which originate near the central
nuclei and move out with relativistic speeds exceeding $0.9\,c$. The jet
plasma radiates strong synchrotron radio emission
which is detectable on parsec-scales by very large baseline
interferometry (VLBI). Here, we consider the flux-density variation of
the jet plasma due to the GR effect, and the possibility of its detection
by present and future radio-astronomical facilities.

\section{Flux variation of the relativistic jet}
Suppose that the jet plasma component radiating a power law spectrum
$F_{\rm e}(\nu) \sim \nu^{-\alpha}$ ($\alpha$ is the spectral index) 
moves in the gravitational field of 
SMBH with a speed $\beta$ in a direction with an angle $\theta$ to
the line of sight of the observer. Then the observed flux at a time
$t$ depends on the Doppler factor $\delta_{\rm t}(\beta; \theta)$ of
the component and the distance of the component from the SMBH at $t$, $R_{\rm t}$,
\begin{equation}
  F_{\rm t}(\nu) = k_{\rm t}^{3+\alpha}\,F_{\rm e}(\nu);\,\,\,\,\,\,k_{\rm t}=\frac{\delta_{\rm
  t}}{1+z}\left ( 1-\frac{R_{\rm s}}{R_{\rm t}} \right )^{1/2};
\end{equation}
where $R_{\rm s}$ and $z$ are the Schwarzschild radius of the SMBH and its redshift.
%
%
The detected flux density may vary as a result of
deceleration or acceleration of the jet component, bending of the jet and
intrinsic or extrinsic variability of radio emission.  To investigate the
pure effect of the GR on the flux density
variation, we assume, that the $F_{\rm e}(\nu)$ and $\delta_{\rm t}$
remain unchangeable in time. At a fixed frequency $\nu_{\rm f}$, the
detected flux density $F(\nu_{\rm f})$ of the jet component grows 
with $R_{\rm t}$ (Fig. 1) reaching up to 90 \% of its intrinsic flux
density at $\sim 20R_{\rm s}$, and then gradually approches to $\sim
100$ \% at $\sim 100R_{\rm s}$.

\section{Prospects of detecting the GR}
Seven millimeter VLBI\footnote{Krichbaum et al., poster \#43 in this
proceedings (see also:
http://www.mpifr-bonn.mpg.de/staff/tkrichbaum/3c274.html)} (43 GHz)
can achieve an angular resolution of $\sim 70\,\mu$as corresponding to
a spatial resolution of $0.0058\,{\rm pc} \simeq 19R_{\rm s}$ for
M87\footnote{M87 (3C 274) has a radio bright core-jet structure, 
$z=0.0044$ ($D \sim 17.1$ Mpc, for a flat cosmology with $H_{\rm
0}=70$ km s$^{-1}$Mpc$^{-1}$), $M_{\rm BH} \sim 3\times10^9\,M_{\odot}$
($R_{\rm s}=0.0003$ pc)}. Full coverage global mm-VLBI imaging with
more sensitive radio telescopes will allow even more detailed and
better quality images$^{1}$. Present high-resolution multi-epoch
mm-VLBI imaging of M87 is capable of detecting $\sim 10$ \% flux
density variations (Fig. 1) from the relativistic jet plasma at $\sim$
(15 to 100)$R_{\rm s}$ due only to GR effect.

At lower frequency, a global VLBI array enhanced by the SKA will allow one to study the nearest
weak AGN which previously were unreachable by VLBI due to their
faintness. The next generation space-VLBI mission
VSOP-2\footnote{Hirabayashi et al., poster \#4 in this proceedings}
will achieve the highest resolution of 38 $\mu$as at 43 GHz, which is
twice as high ($\sim 10R_{\rm s}$ for M87) as the resolution of a
present mm-VLBI. The resolution of a global sub-mm-VLBI$^{1}$ at 230
GHz will be $<30\,\mu$as which is comparable with the resolution of
future space-VLBI missions. The global sub-mm-VLBI combined with ALMA can
achieve the unprecedented angular resolution of $\sim 10\,\mu$as
($\sim 3\,R_{\rm s}$ for M87!) allowing the gravitational effects to
be tested in the vicinity of SMBH thus providing additional test
for the theory of general relativity and a new method for measuring
the masses of SMBH.

TGA is grateful to the AvH Foundation for a
Humboldt Fellowship, and to Drs. T. Krichbaum, A. Lobanov and
A. Polatidis for useful discussions.

\begin{figure}[b]
\begin{center}
\includegraphics[angle=-90,width=.8\textwidth]{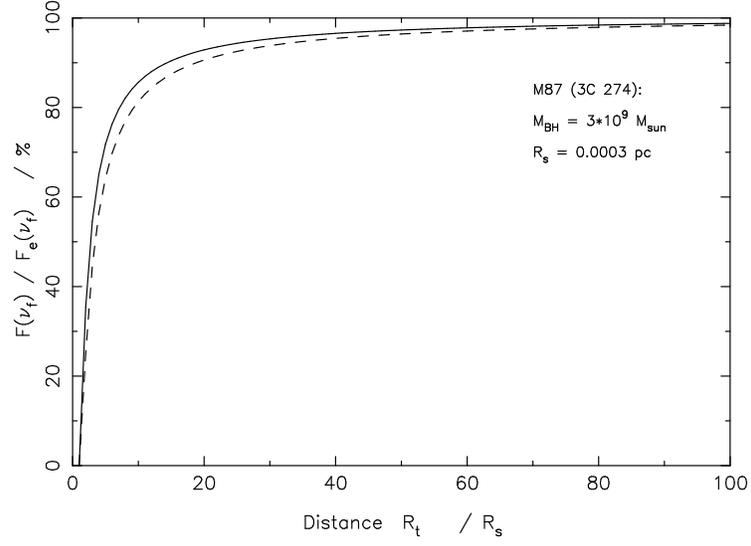}
\end{center}
\caption[]{The observed flux density normalized to the intrinsic flux density of the jet component
versus its distance (in $R_{\rm s}$ units) from the SMBH. Amplification of
the flux density due to GR effect is shown for a  
flat $\alpha=0$ (solid line) and steep $\alpha=1$ (dashed line) spectra.}
\label{eps1}
\end{figure}

\end{document}